\documentclass[12pt]{iopart}

\usepackage{iopams}  
\usepackage{graphicx}

\usepackage{hyperref}
\usepackage{xcolor}
\usepackage{verbatim}
\newcommand{\bea}{\begin{eqnarray}}
\newcommand{\eea}{\end{eqnarray}}

\begin{document}

\title[Random number generation \& distribution out of thin (or thick) air]{Random number generation \& distribution out of thin (or thick) air}

\author{N Bornman$^1$, A Forbes$^1$ and A Kempf$^{2,3}$}

\address{$^1$School of Physics, University of Witwatersrand, Johannesburg 2050, South Africa}
\address{$^2$Department of Applied Mathematics, Department of Physics and Astronomy and Institute for Quantum Computing, University of Waterloo, Waterloo, Ontario, Canada N2L 3G1}
\address{$^3$Perimeter Institute for Theoretical Physics, Waterloo, Ontario, Canada N2L 2Y5}

\ead{nicholas.bornman1@students.wits.ac.za}
\vspace{10pt}
\begin{indented}
\item[]March 2020
\end{indented}

\begin{abstract}
Much scientific work has focused on the generation of random numbers as well as the distribution of said random numbers for use as a cryptographic key. However, emphasis is often placed on one of the two to the exclusion of the other, but both are often simultaneously important. Here we present a simple hybrid free-space link scheme for both the generation and secure distribution of \mbox{(pseudo-)}random numbers between two remote parties, drawing the randomness from the stochastic nature of atmospheric turbulence. The atmosphere is simulated using digital micro-mirror devices for efficient, all-digital control. After outlining one potential algorithm for extracting random numbers based on finding the centre-of-mass (COM) of turbulent beam intensity profiles, the statistics of our experimental COM measurements is studied and found to agree well with the literature. After implementing the scheme in the laboratory, Alice and Bob are able to establish a string of correlated random bits with an 84\% fidelity. Finally, we make a simple modification to the original setup in an attempt to thwart the hacking attempts of an eavesdropper, Eve, who has access to the free-space portion of the link. We find that the fidelity between Eve's key and that of Alice/Bob is 54\%, only slightly above the theoretical minimum. Atmospheric turbulence could hence be leveraged as an added security measure, rather than being seen as a drawback.
\end{abstract}

\submitto{\JOPT}


%
%
%
%
%

\maketitle

\section{Introduction}
\label{sec:introduction}

Contemporary random number (RN) generation protocols are legion and based on a wide variety of processes. In a classical setting, studies which exploit the stochastic nature of atmospheric turbulence \cite{marangon2014random, font2015characterization}, phase and frequency jitters in oscillators comprised of semiconductors and lasers \cite{dichtl2007high, reidler2009ultrahigh, kanter2010optical}, chaotic maps \cite{stojanovski2001chaos} and which even use de-correlated photographs of lava lamps to seed a classical pseudo-random number generator \cite{johansson1999random} have been performed. However, given that classical physics is ultimately deterministic, there is broad consensus that non-deterministic quantum processes result in superior RNs: one could extract randomness from various degrees of freedom for both single and entangled photons in quantum optics setups \cite{furst2010high, fiorentino2007secure, pironio2010random}, radioactive decay in atoms \cite{figotin2004random}, or even from the quantum vacuum \cite{gabriel2010generator}. Despite all the options available, the importance of efficient, true random number generation in science and broader society cannot be overstated: RNs play central roles in numerical solutions to otherwise intractable mathematical problems, Monte-Carlo simulations \cite{metropolis1949monte}, data encryption and secure communication in the form of keys \cite{stallings1999cryptography}, as well as in weather modeling \cite{palmer2005representing}.

In the case of RNs applied to cryptography and secure communication, one party, Alice, would ordinarily generate RNs, and send them (or a derivative of them) to a second party, Bob. They would then use the shared information to encrypt their communication channel. However, a problem with such a scheme is not the inability of Alice to generate RNs: indeed, cost-efficient, compact random number generators exist on the market. A potential weakness rather is the distribution of the RNs to Bob, in a remote location, for use as an encryption key: this distribution is open to being intercepted by an eavesdropper, Eve. Many modern schemes, such as public-key cryptography, use a combination of different keys and computational complexity to combat this \cite{stallings1999cryptography}. An ideal secure communication protocol, however, would entail both Alice and Bob remotely and independently deriving the same strings of random numbers for use as keys, while being sure that an eavesdropper is unable to derive the same numbers. Given shared keys, schemes such as one-time-pads \cite{stallings1999cryptography} then guarantee security.

Quantum protocols exist which ensure secure communication between Alice and Bob by both preventing Eve from gleaning sufficient information about the key and alerting the participants to Eve's presence. However, quantum protocols are often highly unwiedly and impractical. Furthermore, some ultimately deterministic but stochastic processes do possess enough randomness to prove useful.

The purpose of the current study is to investigate the potential of a communication link using one such process, namely atmospheric turbulence, to both generate and distribute the same set of (pseudo-)random numbers securely between two parties, while preventing a malevolent third party from accessing the RNs. The authors are only aware of two somewhat related studies which have considered this possibility before \cite{thomas2015phase, drake2013optical}. However, these studies used phase information to generate RNs, and consisted of relatively complex setups using interferometers and multiplexers. Our proposed setup is far simpler (using only beamsplitters and CCD cameras), and with further work, may well match the security and efficiency of other protocols.

\begin{figure*}
\centering
\includegraphics[width=\linewidth]{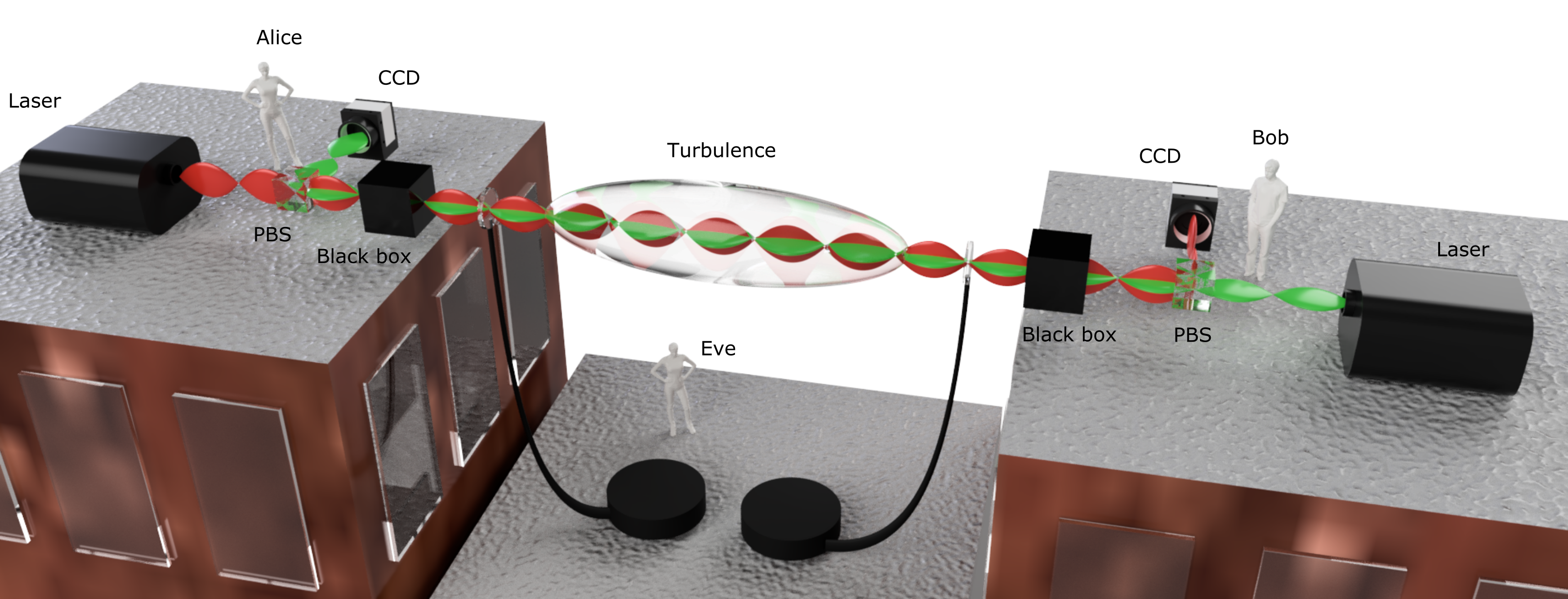}
\caption{Schematic of proposed hybrid random number generation and distribution free-space link between Alice and Bob, with an attacker Eve able to hack the free-space portion of the channel.}
\label{fig:schematic}
\end{figure*}

Fig. \ref{fig:schematic} is a simple schematic of the proposed setup. Alice and Bob, in two remote locations, create a free-space link comprised of two co-linear but counter-propagating laser beams with orthogonal polarisations. Alice's beam propagates, through atmospheric turbulence, to Bob's terminal of the link where it is diverted to a CCD camera (under Bob's control) by a polarising beamsplitter. Bob's counter-propagating beam undergoes the same steps, with Alice measuring his perturbed beam. Since atmospheric turbulence changes relatively slowly (on the order of a millisecond \cite{schmidt2010numerical}), both parties measure each other's beams which are ultimately correlated since they each encounter the same atmospheric distortion. These correlated beam intensity profiles can then be used by Alice and Bob to remotely and independently extract correlated random numbers for use as an encryption key. In this way, the two parties can actually derive benefit from atmospheric turbulence, which is usually the bane of optical free-space links.

At this stage, it is technically possible for Eve to hack the system if she were to, for example, place beamsplitters at either terminal of the link. However, a simple addition could thwart her efforts: Alice and Bob each simply insert separate black boxes at their respective ends of the link, before the free-space portion (see figure \ref{fig:schematic}), which further randomly perturb both co-linear beams. If the initial beam profiles are kept secret (so that only Alice has knowledge of her initial beam profile, and Bob his) and Eve also has no direct access to the black boxes (i.e. she can only hack the free-space portion of the link), any beam Eve observes will not have undergone the same combination of distortions as the beams Alice and Bob measure. Hence, Eve cannot extract the same RNs, even with full knowledge of the algorithm Alice and Bob use to extract the RNs.

The outline of this paper is as follows. Section \ref{sec:theory} outlines the theory needed to understand contemporary studies of atmospheric turbulence. Section \ref{sec:experiment} describes the experimental setup; section \ref{sec:turbulence} gives an explanation of how to simulate turbulence using both spatial light modulators (SLMs) and digital micro-mirror devices (DMDs), while section \ref{sec:rngscheme} outlines the algorithm used to extract RNs from the turbulent beam intensity profiles. Finally, section \ref{sec:resultsanddiscussion} discusses in detail the obtained results, with conclusions presented in section \ref{sec:conclusion}.

\section{Theoretical Background}
\label{sec:theory}

Understanding turbulence in fluids is an incredibly intricate problem: the Clay Mathematics Institute holds a full understanding of the Navier-Stokes equations, which fully govern viscous fluid flow, in such high regard that a solution will earn the solver $\$ 1 $million. Here, we give a brief summary of salient points in modern simulations of beam propagation through turbulence.

Atmospheric turbulence is a stochastic process which causes random spatial and temporal fluctuations in the refractive index of the medium, which in turn causes variations in the optical path, intensity and phase of laser light traveling through it. For paraxial laser light (of wavenumber $k_0$ in the vacuum) propagating in the $z$ direction and emanating from a source at the plane $z = 0$, the random phase change $\theta(\mathbf{X})$ due to turbulence is

\bea
\fl \theta(\mathbf{X}) = k_0 \int_0^L \delta n(\mathbf{x}) dz,
\label{eqn:phasechange}
\eea

\noindent
where $\delta n(\mathbf{x})$ is the refractive index fluctuation at point $\mathbf{x} = (x,y,z)$, $\mathbf{X} = (x,y)$, and $z = L$ is the distance the light has propagated in the $z$ direction. Understanding $\delta n$ is hence central to understanding turbulence. However, $\delta n$ is chaotic and hence stochastic methods are required. As such, information about the refractive index needs to be gleaned from correlation functions. In our case the structure function of the index of refraction fluctuation, $D_n(\mathbf{x}_1,\mathbf{x}_2)$, between two points in space is important

\bea
\fl D_n(\mathbf{x}_1,\mathbf{x}_2) = \langle \left[ \delta n(\mathbf{x}_1) - \delta n(\mathbf{x}_2) \right]^2 \rangle.
\label{eqn:structurefunction}
\eea

In 1941, A. N. Kolmogorov modeled turbulence as consisting of small, randomly varying eddies of constant pressure and temperature, which exchange energy among themselves. For eddy sizes larger than an `inner' scale $l_0$ and smaller than an `outer' scale $L_0$, Kolmogorov argues that the eddies are homogenous and isotropic \cite{schmidt2010numerical}. Within this subrange, $D_n$ is given by his famous `two-thirds power' law

\bea
\fl D_n(r) = C^2_n r^{2/3},
\label{eqn:twothirds}
\eea

\noindent
where $\mathbf{r} = \mathbf{x}_2 - \mathbf{x}_1$, $r = |\mathbf{r}|$ and $C_n^2$, the refractive index structure constant, is a small scalar characterising the strength of the turbulence. The refractive index auto-correlation function is related to $D_n$ by \cite{ibrahim2013orbital}

\bea
\fl \langle \delta n(\mathbf{x}_1)\delta n(\mathbf{x}_2) \rangle = \langle \delta^2 n(0) \rangle - \frac{1}{2}D_n(r).
\label{eqn:autocorrelation}
\eea

Note that the homogeneity and isotropy assumptions imply that the auto-correlation function depends only on the difference between the co-ordinates, so $\langle \delta n(\mathbf{x}_1)\delta n(\mathbf{x}_2) \rangle = \langle \delta n(0) \delta n(\mathbf{x}_2 - \mathbf{x}_1) \rangle$. Furthermore, the \textit{Wiener-Khinchin} theorem states that the power spectral density (PSD) of the refractive index fluctuation, $\Phi_n(\mathbf{k})$, which gives a measure of the statistical distribution of the abundance and size of the turbulent eddies, is the Fourier transform of this auto-correlation function,

\bea
\fl \Phi_n(\mathbf{k}) = \mathcal{F}\{\langle \delta n(0) \delta n(\mathbf{x}) \rangle \}(\mathbf{k}),
\label{eqn:wiener-khinchin}
\eea

\noindent
where $\mathbf{k}$ is the 3-dimensional wavevector. If we assume the turbulence to be Markovian, then $\mathbf{k}_z = 0$. Finally, combining equations (\ref{eqn:twothirds}) to (\ref{eqn:wiener-khinchin}), a simple expression for the Kolmogorov refractive index PSD results

\bea
\fl \Phi_n(\mathbf{k}) = 0.033 C_n^2 |\mathbf{k}|^{-11/3}.
\label{eqn:ripsd}
\eea

A more in-depth analysis would detail the derivation of an expression for the phase change due to turbulence, equation (\ref{eqn:phasechange}), in terms of a random normally distributed complex spectral function. However, we shall simulate turbulence numerically, drawing samples from a distribution outlined in section \ref{sec:turbulence}. See \cite{ibrahim2013orbital, phillips2005atmospheric} for more detailed analyses.

\section{Experimental Setup}
\label{sec:experiment}

\begin{figure*}
\centering
\includegraphics[width=0.8\textwidth]{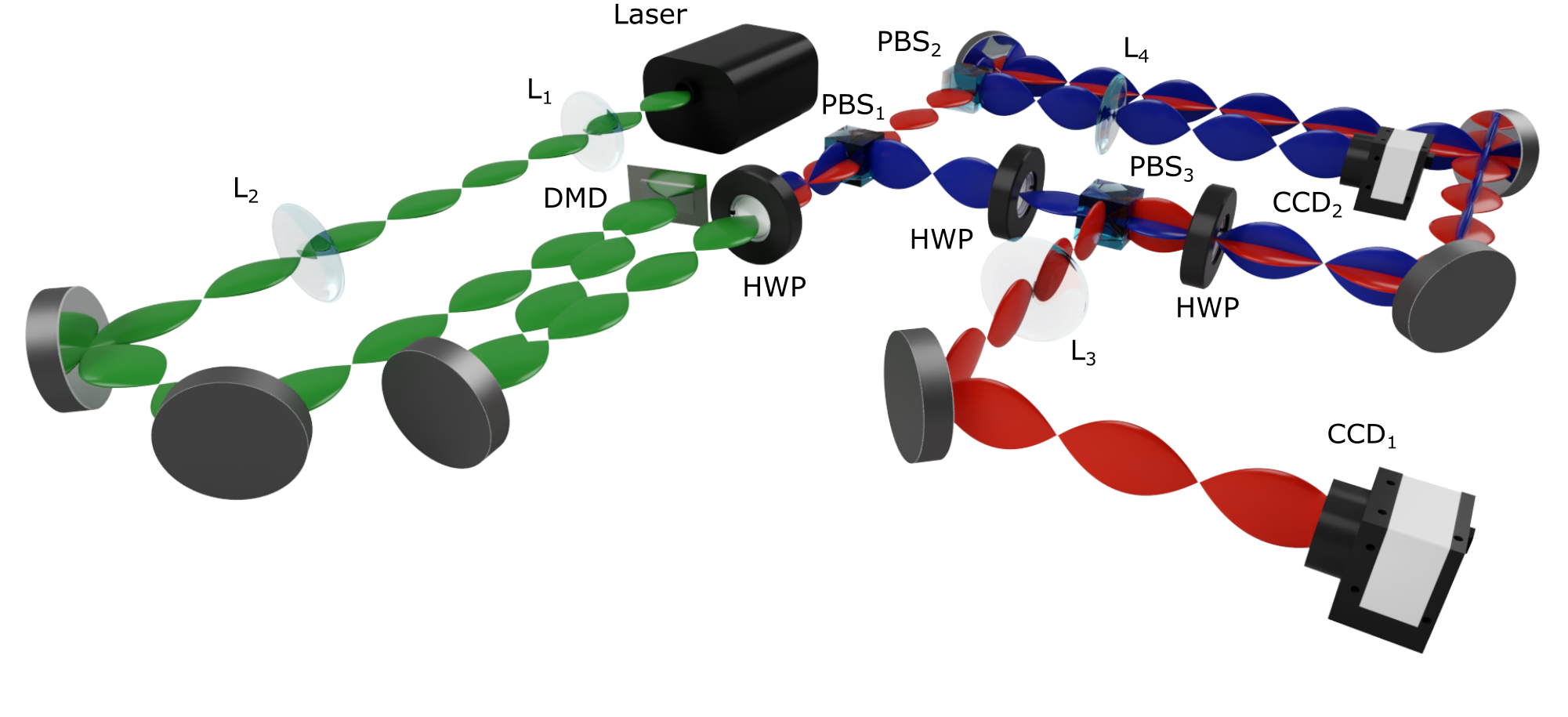}
\caption{After collimating the beam with lenses L1 (f = 50mm) and L2 (f = 150mm), a DMD modulates the photons, a half-wave plate (HWP) creates an equal superposition of horizontally- and vertically-polarised light, and a polarising beamsplitter (PBS) splits the light into these two polarisations. A sequence of HWPs and PBSs then create a loop, with the latter diverting vertically-polarised light onto lenses L3 and L4 (f = 750mm) followed by CCD cameras. Note that the wavelength of the light is 514nm throughout the experiment: the red (transmitted from PBS1) and blue (reflected) colour of the pulses simply indicate the paths traveled (colour online).}
\label{fig:setup}
\end{figure*}

The aim of the current paper is to present proof-of-principle evidence for a free-space link which acts as a dual (pseudo)-random number generation (RNG) and distribution scheme in a classical setting. As such, the experimental setup is shown in figure \ref{fig:setup}. A 514nm Gaussian laser beam was expanded to a waist of 1mm and collimated using two lenses (L1, f = 50mm and L2, f = 150mm). This light was directed onto a DLP3000 DMD, initially masked with just a grating for alignment purposes. Thereafter, the first order from the DMD, passing through a half-wave plate, resulted in equal intensities of horizontally- and vertically-polarised light. A 50:50 polarising beamsplitter (PBS1) split the beam into two and mirrors were used to create a closed loop.

Directly after the transmitting output port of PBS1, an identical PBS (PBS2) was placed which transmitted horizontally-polarised light and reflected the vertical component; after the reflecting output port, a half-wave plate, a third PBS (PBS3), and another half-wave plate were placed. Mirrors directed the beam from this last half-wave plate to PBS2, closing the loop. Horizontally-polarised light transmitted through PBS1 passed unhindered through PBS2, but was reflected by PBS3; vertically-polarised light reflected by PBS1 passed unhindered through PBS3, but was reflected by PBS2. A 750mm lens (L3 and L4) was placed in each of the output ports of PBS2 and 3, along with two CCD cameras to image the Fourier plane of the DMD. In such a setting, both beams experienced the same turbulence modulation since the original beam was modulated with the DMD before being split into polarisation components (in practice, one would need to ensure that both independent beams are roughly the same size so that they experience the same aberrations). This splitting of the modulated beam, and the closed loop, serve to model co-linear, counter-propagating beams Alice and Bob would employ in practice. It`s worth mentioning that it is not strictly necessary to create the closed loop, and a spatial light modulator (SLM), another popular device for effecting changes to a beam's profile, could have been used in place of the DMD. However, if one wished to modulate the beams inside the loop, a DMD is necessary since SLMs don't modulate vertically-polarised light. Unfortunately, we found that modulating the beams with the DMD inside the loop distorted the one beam with respect to the other far too much to be of practical use. This could be due to imperfections in the way the DMD modulates the beams for different incident angles and directions: such incident angles are necessary if counter-propagating beams are to be modulated inside the loop.

\section{Turbulence Simulation Procedure}
\label{sec:turbulence}

Atmospheric turbulence aberrations are a phase effect, and are hence described by a phase-only transfer function $T(x,y) = e^{i\phi(x,y)}$. Here, $\phi$ is the phase change in equation (\ref{eqn:phasechange}) and $(x,y)$ are the transverse co-ordinates in the plane $z = L$, after the beam has propagated a distance $L$ through turbulence. Computer-generated phase masks representing this turbulence transfer function are generated by first drawing random samples from a specific statistical distribution and transforming these samples into a 2-dimensional grid of phase values. The distribution matches the statistics of turbulence so that the phase values of the grid mimic the statistics of turbulence-induced phase perturbations. Strictly speaking, the phase variation along the optical path of the beam is due to many random factors which, individually, potentially don't follow a well-defined distribution. However, by the central limit theorem from probability theory, the average over all of these small contributions is in fact well approximated by a normal distribution. Hence, we can numerically generate the grid of turbulence-induced phase changes, i.e. $\phi$, using the procedure outlined in \cite{schmidt2010numerical, toselli2015slm}. First we create an array of $N_x = 608$ (the number of pixels in the horizontal direction of our DMD) equally-spaced points in the range $\left[ -\frac{N_x - 1}{2 N_x \Delta x} : \frac{N_x - 1}{2 N_x \Delta x} \right]$, where $\Delta x = 10.8 \mu m$ is the size of the pixel in the $x$ direction. Similarly, for the $y$ direction, $N_y = 684$ and $\Delta y = 5.4 \mu m$. These two arrays together form a 2D grid of ordinary spatial frequency co-ordinates, $(f_x, f_y)$. Next, the Kolmogorov refractive index PSD, equation (\ref{eqn:ripsd}), is related to the phase PSD, $\Phi_{\phi}(\mathbf{k})$, in the transverse plane via

\bea
\fl \Phi_{\phi}(\mathbf{K}) = 2 \pi^2 k_0^2 L \Phi_n(\mathbf{K}),
\label{phasepsd}
\eea

\noindent
where $\mathbf{K} = (2\pi f_x,2\pi f_y)$ ($k_z = 0$ under the assumption of a Markovian process and the single phase screen approximation). Note that if one wished to use a different PSD to model turbulence, such as the von K\'arm\'an or modified von K\'arm\'an PSDs \cite{schmidt2010numerical}, one need simply replace $\Phi_n$ in equation (\ref{phasepsd}) with the appropriate model. Equation (\ref{phasepsd}) gives the covariance of the turbulence's normal distribution at point $\mathbf{K}$ (with a mean of zero). The root of equation (\ref{phasepsd}) is evaluated on the above frequency grid, with the result multiplied element-wise by a second grid of random, normally-distributed complex numbers with both the real and imaginary parts having mean zero and standard deviation of one. Finally, taking the inverse Fast Fourier Transform of the result gives $\phi$.

\begin{figure}[t]
\centering
\includegraphics[width=\linewidth]{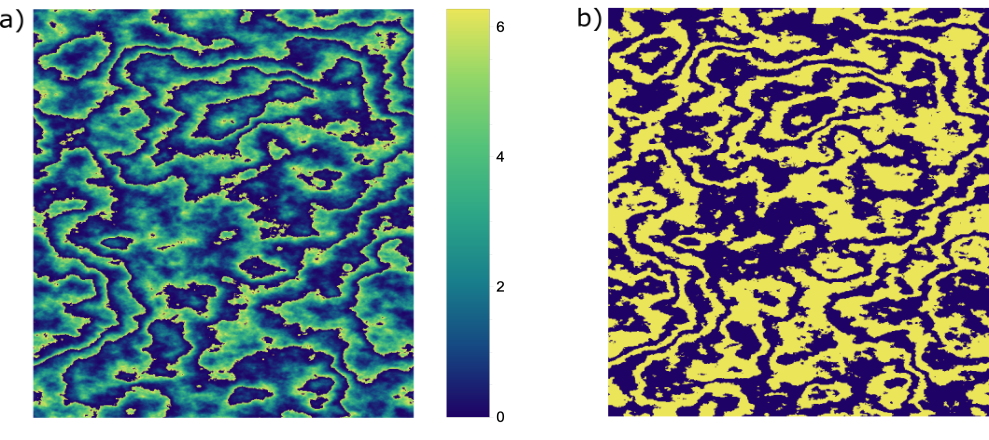}
\caption{Numerically-simulated turbulence mask example. a) initial phase mask (phase shown in the legend), b) binarised version (with blue (yellow) corresponding to the off (on) mirror position). Here $k_0 = 2 \pi / (514nm)$, $L = 10km$ and $C_n^2 = 10^{-15} m^{-2/3}$.}
\label{fig:masks}
\end{figure}

The transfer function $T(x,y)$ could now be computed using the grid of $\phi$ values and masked directly onto a spatial light modulator (SLM) to simulate turbulence (a procedure that has been well studied in the literature \cite{burger2008simulating}). In the case of a DMD, however, one last step is necessary: as an amplitude-only device consisting of an array of pivoting micro-mirrors, each pixel of a DMD can only adopt one of two positions: `on' or `off'. As such, the $\phi$ grid of any hologram first needs to be `binarized' to either 0 or 1, representing the `off' and `on' pixel state respectively. Although a DMD can be employed in experiments requiring complex amplitude modulation \cite{Lee:79, anzuola2016generation}, as mentioned, atmospheric turbulence is a phase-only phenomenon. Therefore, the transfer function $T(x,y)$ is simulated on a DMD with a hologram found according to \cite{anzuola2016generation}

\bea
\fl h(x,y) = \frac{1}{2} + \frac{1}{2}\mbox{sign}(\cos(2\pi \alpha + \phi(x,y))),
\label{eqn:dmdeqn}
\eea

\noindent
with $\alpha$ a diffraction grating (recall that we image the first order from the DMD). Figure \ref{fig:masks} gives an example of a simulated turbulence mask.

\section{Random Number Generation Scheme}
\label{sec:rngscheme}

\begin{figure}[t]
\centering
\includegraphics[width=\linewidth]{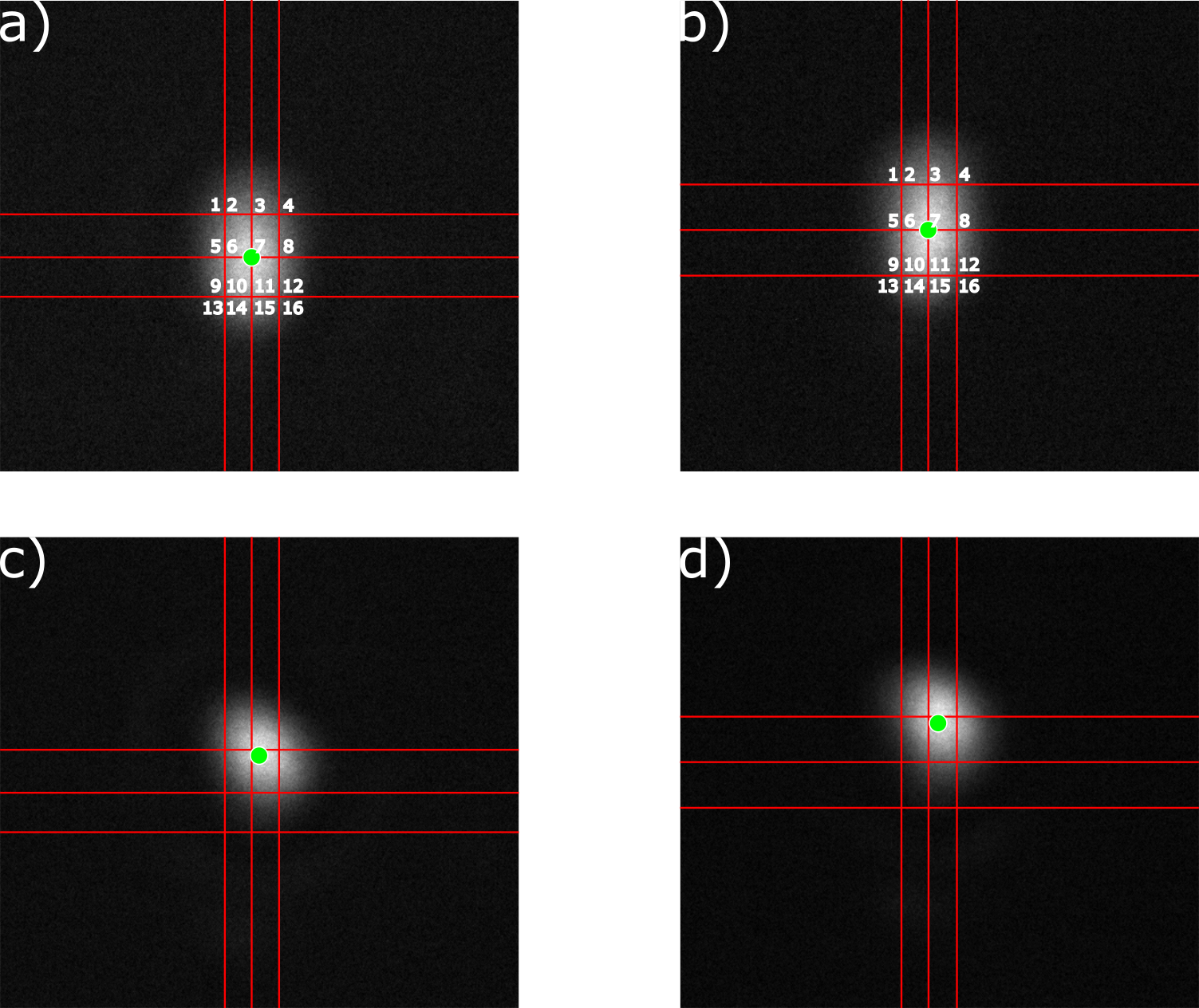}
\caption{a) and b) give intensity profiles for Alice and Bob's unperturbed beams, overlayed by horizontal and vertical lines separating the four quartiles in both the $x$ and $y$ direction, respectively. The beam COM is marked with a green dot, and the inlaid digits indicate the 4-bit number associated with each rectangle. c) and d) give the respective profiles for a single turbulent frame. The COM in each case lies in rectangle 7, so 7 is the random number Alice and Bob each assign to this particular intensity profile.}
\label{fig:data}
\end{figure}

Here we outline, given a set of $N$ pre-generated turbulence masks, how Alice and Bob can each individually generate a correlated string of random bits from the $N$ intensity profiles they measure, as per section \ref{sec:experiment}. Data for a single, representative run is given in figure \ref{fig:data}. An evident spatial correlation exists between the turbulent profiles in c) and d), hence it is possible for Alice and Bob to extract a similar RN from each image they individually capture. The source of the randomness of the extracted numbers is the (simulated) atmospheric turbulence. Furthermore, since the two co-linear but counter-propagating beams experience the same turbulence fluctuations if the data capturing is synchronised between the two parties and is fast enough, correlations will exist between the RNs Alice and Bob generate if they use the same algorithm to extract said RNs. It is possible to conceive of a number of such extraction algorithms with varying degrees of accuracy and efficiency. We used a simple post-processing scheme: after Alice and Bob have each remotely captured their $N$ images, they find the centre-of-mass (COM) of each. As argued in section \ref{sec:turbulence} and as will be seen in section \ref{subsec:RNG}, the $x$ and $y$ co-ordinates of the COM results follow separate normal distributions. Alice and Bob individually find the standard deviations of their lists of $x$ and $y$ COM co-ordinates. With this, they can generate a random number based on the quartile a particular intensity profile's COM co-ordinate lies in (see figure \ref{fig:data}); more often than not, they will arrive at the same number.

For example, after collecting all the data, suppose that Alice's COM $x$ co-ordinate standard deviation is $1$ and Bob's is $0.8$ (we assume without loss of generality that the mean of both lists is $0$). The quartile 3 range for Alice's $x$ list is hence $[0,0.67]$, and Bob's quartile 3 range is $[0,0.61]$. For a single turbulent image profile, suppose that Alice calculates a COM $x$ value of $0.31$ while Bob calculates a value of $0.28$ for the corresponding image he captures. Both of these values lie in their respective quartile 3 ranges and so Alice and Bob each assign the number 3 to that image (an identical procedure is repeated for the $y$ co-ordinate, doubling the bits extracted from each image). Repeating this for each image, Alice and Bob can generate a string of correlated bits with each image giving four bits (different RN extraction algorithms may potentially give more bits per image).

Note that if an individual COM co-ordinate measurement lies close to the `edge' of the quartile range, Alice and Bob may assign different random numbers to that measurement (see figure \ref{fig:data} c) and d), where the COM lies close to the edge of the rectangle). Such inevitable disparities between the bits Alice and Bob eventually extract can easily be both detected and rectified using contemporary classical error-correcting protocols from coding theory \cite{macwilliams1977theory}, at the expense of bit rate.

\section{Results and Discussion}
\label{sec:resultsanddiscussion}

\subsection{Random Number Generation}
\label{subsec:RNG}

We first confirmed the statistics of the experiment's COM measurements from a sample of images (which should theoretically obey equations (\ref{eqn:ripsd}) and (\ref{phasepsd})), given that it is perhaps not initially evident that the measurements follow the turbulence's statistics. Since $\Phi_{\phi}$ gives the covariance of turbulence and $\Phi_n \propto C_n^2$ for the Kolmogorov model PSD, a log-log plot of the standard deviation (of data derived from a process which should possess a covariance given by $\Phi_{\phi}$) versus $C_n^2$ should yield a linear relationship. This was indeed the case: as the strength of the turbulence increased (by increasing the $C_n^2$ value of the mask), the spread of the turbulent beam's COM increased too. This linearity wouldn't quite hold true for strong turbulences, however: since tip and tilt are the dominant aberrations caused by atmospheric turbulence (especially in the weak irradiance regime \cite{roggemann2018imaging}), the perturbed beams have mostly only had their transverse positions, and hence COMs, shifted with respect to the unperturbed beams. This would generally be the case regardless of the initial beam profile (within certain limits on the overall beam waists). However, other types of atmospheric aberrations such as astigmatism become more pronounced in stronger turbulence and would not be compatible with the algorithm proposed. Despite this, a real-world application could easily effect a cutoff during such scenarios. Stronger turbulence in free-space quantum key distribution schemes tends to have deleterious effects on the schemes' bit rates and overall robustness \cite{goyal2016effect}. It should be possible, on the other hand, to devise an alternative RN extraction algorithm which exploits stronger turbulence situations in the current proposed scheme, potentially turning a negative into a positive.

\begin{figure}
\centering
\includegraphics[width=\linewidth]{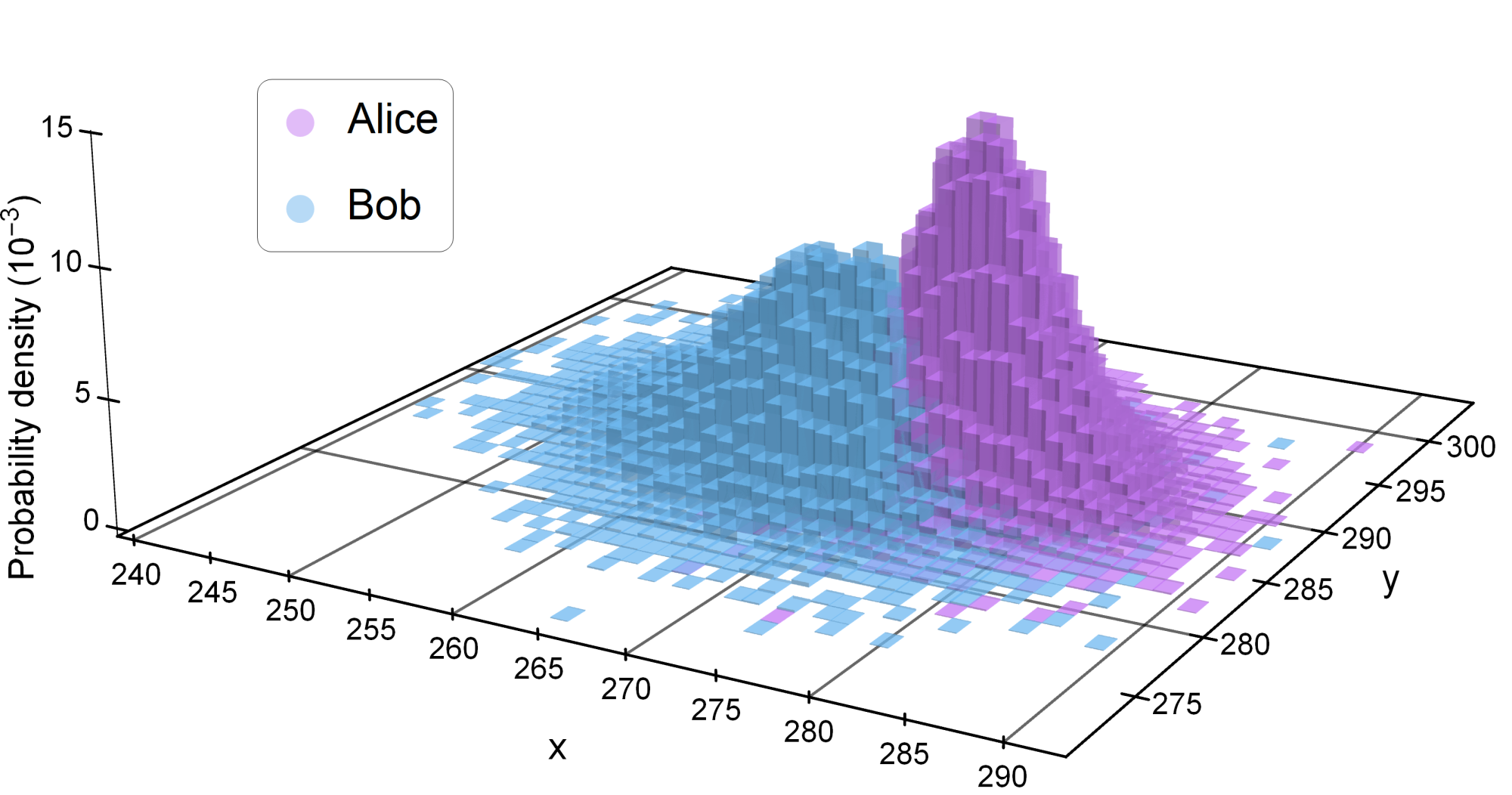}
\caption{Probability distribution of Alice and Bob's centre-of-mass measurements for 30,000 turbulent profiles, as a function of the cameras' $x$ and $y$ pixel co-ordinates.}
\label{fig:histogram3D}
\end{figure}

\begin{figure*}[t]
\centering
\includegraphics[width=0.9\textwidth]{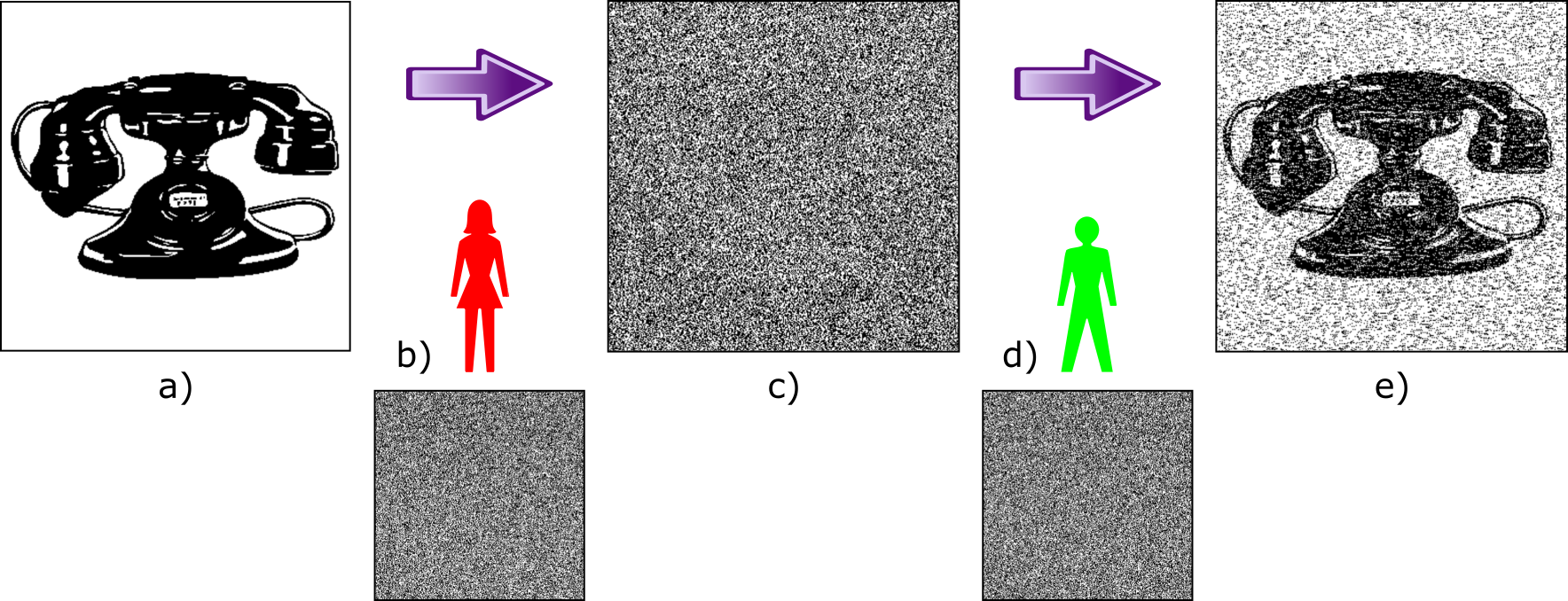}
\caption{Alice, using her RN key (b), encrypts the original binary message (a) (using a simple XOR operation on each pixel), and sends the result (c) to Bob, who subsequently uses his key (d) to decrypt the message, resulting in (e).}
\label{fig:protocol}
\end{figure*}

Next, to test the scheme from start to finish, a set of 30,000 simulated turbulence masks with \mbox{$C_n^2 = 10^{-16}m^{-2/3}, L = 10km$} was generated. For each mask, Alice and Bob measured the corresponding turbulent beam intensity profiles (see figure \ref{fig:data} for a single run). Thereafter, they separately calculated the COM of each image. A histogram of this data is shown in figure \ref{fig:histogram3D} which clearly follows the anticipated normal distributions. The standard deviations of this data were then used to assign each individual COM $x$ and $y$ co-ordinate to one of four quartiles: values in quartile 1 were assigned a value of 1, quartile 2 a value of 2, etc. The resultant string of 120,000 bits can then be used as a symmetric key to encrypt communication. Figure \ref{fig:protocol} shows the experimentally-generated keys, along with the original message, its encrypted form, and its decrypted form. Visually, the original message is indecipherable from the encrypted image.

It is worth noting that such a system in practice would require some exchange of information during the setting-up of the link, as well as during the initial authentication of Alice and Bob, external to the system itself, to ensure that Alice and Bob are indeed communicating with each other and haven't inadvertently established a link with Eve. This is often the case with commercial encryption devices, and once authenticated, Alice and Bob could in fact incorporate their strings of bits into a continuous authentication protocol.

The decrypted message, figure \ref{fig:protocol} e), appears to possess salt-and-pepper-like noise as a result of the inevitable bit errors between Alice and Bob's keys. However, the quantum of errors is small enough that the original message is still clearly evident and the final message could easily be cleaned up using commonplace image processing software. In particular, the ratio of $0$'s to $1$'s for Alice and Bob's keys is close to $50\%$: $50.55\%$ and $50.10\%$, respectively. It is obviously crucial that these ratios be as close to $50\%$ as possible: for other RN schemes which don't maintain this ratio, we unsurprisingly found that one could visually make out the original message in the encrypted image. Furthermore, the fidelity between Alice and Bob's 120,000-bit keys was found to be $84\%$. Any discrepancy can largely be ascribed to the experimental instability inherent to our simple implementation: slight misalignment over time between both arms will increase the likelihood of a COM measurement being assigned to one quartile in Alice's arm, while being assigned to a different quartile in Bob's, resulting in bit errors for that measurement. However, this issue could simply be overcome by the use of adaptive optics which are widely found in mainstream optical instruments. The use of such optics would even allow for the real-time generation of random numbers, forgoing the post-processing carried out here. Furthermore, although being relatively cheap in comparison to SLMs, the DMD is less efficient and more prone to imperfections (such as thermal heating) during operation, which would also influence the fidelity.

\begin{figure}[t]
\centering
\includegraphics[width=\linewidth]{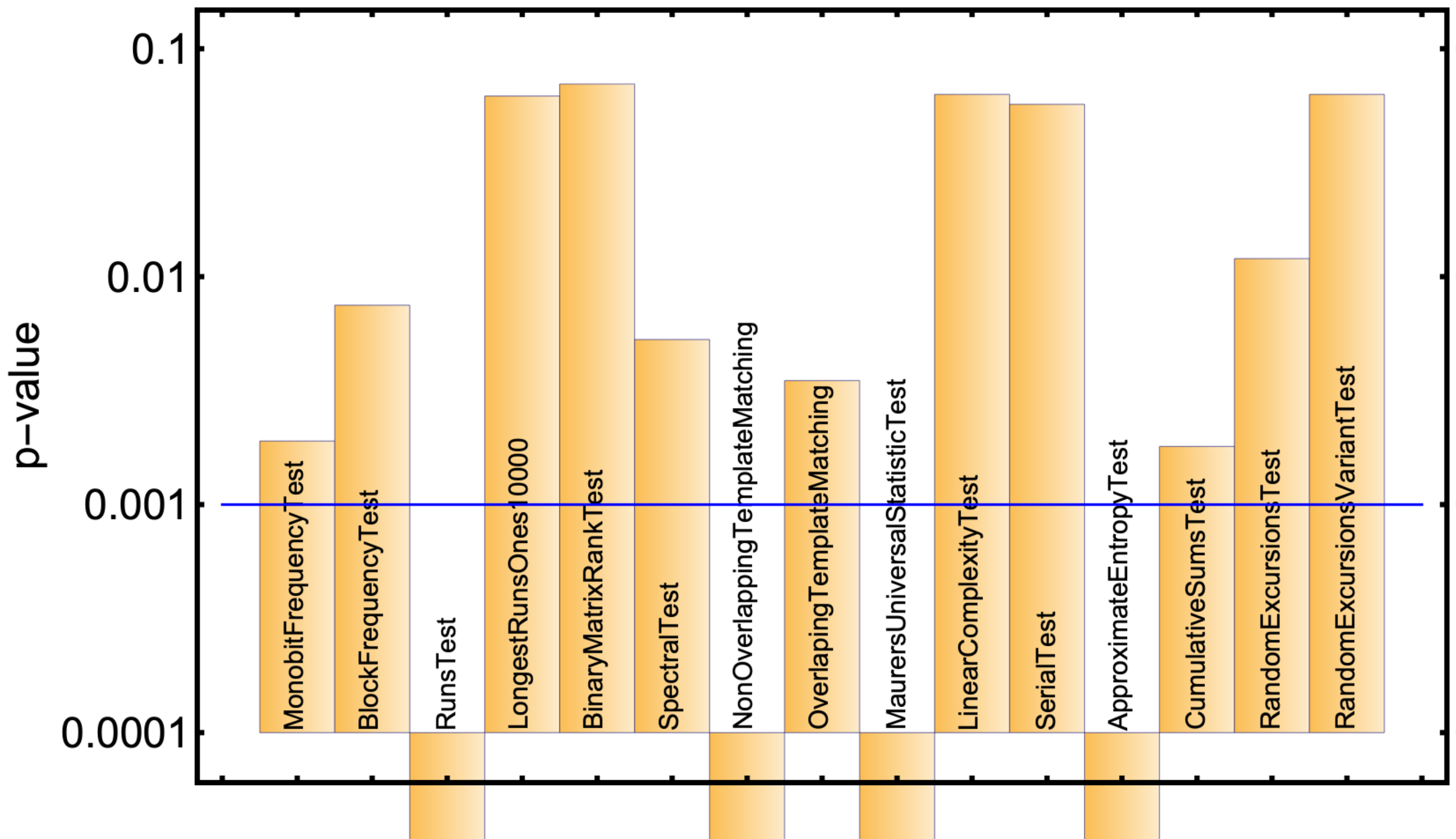}
\caption{Results of the NIST randomness tests applied to Alice's string of bits, giving the p-value corresponding to each test. The horizontal line corresponds to the chosen significance level.}
\label{fig:NIST}
\end{figure}

The randomness of Alice's bits were also tested against the \textit{NIST} random number test suite, a common RNG evaluation standard \cite{NIST}, the results of which are given in figure \ref{fig:NIST}. Alice's bits passed 11 out of the battery of 15 tests, demonstrating the potential of this scheme. Furthermore, the imperfect results are to be expected: Documentation of various randomness tests highly recommends running the tests on large sets of independent results with strings of $10^6$ bits each before being able to draw definitive conclusions as to the protocol's randomness. The simulated turbulence masks were also generated computationally using a pseudo-random number generator and are therefore themselves not truly random (and no strictly deterministic set of subsequent protocols could possibly increase the randomness of the results). This would not present a problem in practice, but further testing is required nonetheless to fully assess the protocol's randomness.

\subsection{Hacking}
\label{subsec:hacking}

Consider the setup of figure \ref{fig:schematic}, less the two black boxes (i.e. that discussed in the previous subsection). Although this scheme is interesting in that two parties can remotely generate the same random number from a mutual source of randomness, such a scheme is vulnerable to hacking were a potential eavesdropper, Eve, to place a beamsplitter at either terminal of the free-space portion of the link and measure the diverted beams. In doing so, she could fully characterise the turbulence and hence the shared key generated by Alice and Bob (assuming she has knowledge of the RN extraction algorithm). However, with two modifications, the setup could frustrate Eve's hacking attempts. 

\begin{figure}[b]
\centering
\includegraphics[width=\linewidth]{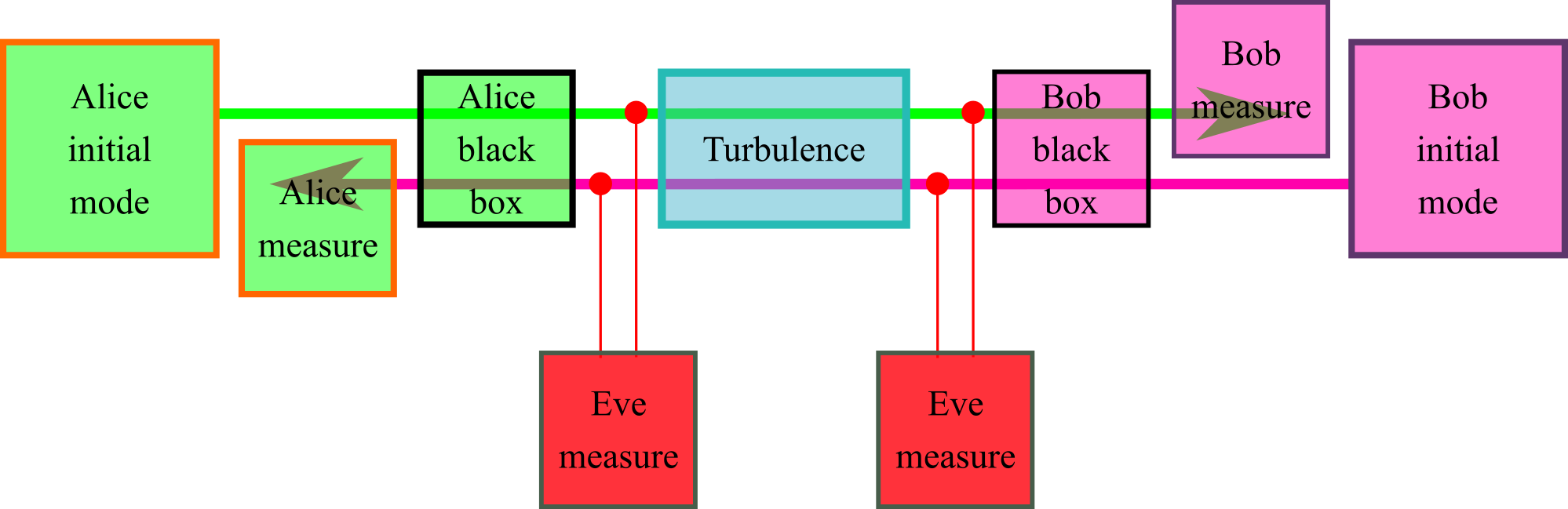}
\caption{Modified setup outlining the addition of the black boxes, and nodes which Eve has access to.}
\label{fig:hack_schematic}
\end{figure}

Firstly, suppose that Alice and Bob were to each add independent black boxes directly at their respective terminals of the link, which randomly perturb both beams at the same rate as the atmospheric turbulence and which Eve does not have access to. For instance, Alice and Bob could publicly declare that they will use black boxes which induce tip and tilt aberrations to the beams (applying the same aberration to both counter-propagating beams), the exact strengths of which are governed by small true random number generators fitted inside each box. In this instance, the randomness for the key is derived from the black boxes and not atmospheric turbulence, which acts merely as a communication link. Secondly, assume that Alice and Bob each independently and secretly shape their initial pump beam profiles to such a degree that Eve cannot characterise the initial beam profiles in real time by measuring the beams emanating from either black box. This situation is presented in the schematic of figure \ref{fig:hack_schematic}.

In such a situation, even if Eve knows the general operation of each black box (e.g. that they effect tip and tilt distortions), with only access to the free-space portion of the link (see figure \ref{fig:hack_schematic}) she can only measure beams perturbed by either the black-box-atmosphere or atmosphere-black-box combination. The intensity profiles Alice and Bob measure have both been perturbed by the black-box-atmosphere-black-box combination. Given this difference, even with knowledge of the RN scheme, Eve cannot easily arrive at exactly the same keys as the authenticated parties in real time.

However, after a certain number of measurements, Eve would presumably be able to deduce the initial pump profiles chosen by both Alice and Bob by comparing successive `initial mode-black box' measurements, since the black box perturbations would be changing with each measurement, not the initial profile. By taking appropriate differences between measurements, she could characterise each beam profile and hence completely hack the system. However, she cannot do this hacking in real time, and Alice and Bob could hence agree beforehand to change their pump profiles regularly and simultaneously. Eve would then have to regularly deduce the new profiles, by which time the authenticated parties could, for instance, have communicated using some sort of time-sensitive self-destructing message. Further work may well consider adding more sophistication to the scheme to further frustrate an eavesdropper.

\begin{figure*}
\centering
\includegraphics[width=0.9\textwidth]{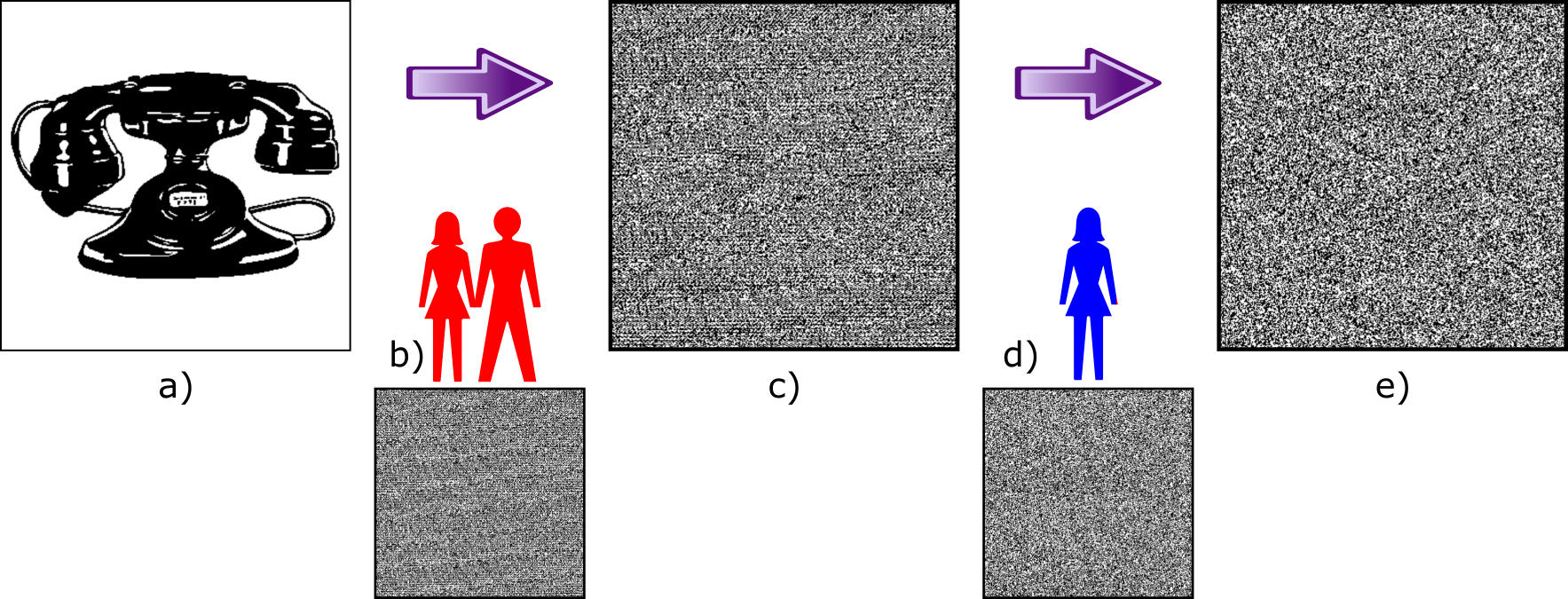}
\caption{Compared with figure \ref{fig:protocol}, Alice/Bob's keys are sufficiently different from Eve's that Eve cannot recover the original message.}
\label{fig:protocol_tp}
\end{figure*}

To investigate experimentally the effect of adding a black box, consider figure \ref{fig:setup} once more. The setup was modified by adding a slowly rotating turbulence plate between PBS1 and PBS2, which acted as a black box. Assume that Alice and Bob measure the same intensity profile as seen by the red beam at CCD1 and that Eve measures the blue beam seen by CCD2 (which only sees the aberrations from the DMD, not the black box). A real-world implementation would include two black boxes, however both Alice and Bob's beams would experience the same perturbations as they propagate through both. With this modification, repeating the exact same procedure as in section \ref{subsec:RNG} gives the results in figure \ref{fig:protocol_tp}. The ratio of $0$'s and $1$'s for Alice/Bob's key and Eve's is $50.01\%$ and $48.88\%$, respectively. However, the fidelity between the keys is $54\%$, only slightly above the theoretical minimum of $50\%$. The addition of the black box appears sufficient to protect the secrecy of the key distribution process.

\section{Conclusion}
\label{sec:conclusion}

In this paper we introduced a simple, proof-of-principle optical free-space link consisting of two co-linear, counter-propagating beams for the generation and distribution of (pseudo-)random numbers, using atmospheric turbulence as the source of randomness. After outlining some salient points underpinning contemporary studies of turbulence, we described the simple experimental setup in which the atmosphere was modeled using a digital micro-mirror device and the two counter-propagating beams modeled by a closed loop. Atmospheric turbulence is known to be largely non-birefringent, so simple polarising beamsplitters can be used to separate the beams at the terminals of the link. Furthermore, the collected data is beam intensity profiles measured by simple CCD cameras instead of phase information required by the more complex experiments of \cite{thomas2015phase, drake2013optical}. The simplicity of our protocol is one such advantage.

After summarising how to simulate atmospheric turbulence with both SLMs and DMDs, we gave an outline of our random number extraction algorithm which uses turbulent beams' intensity profiles' centres-of-mass. Later work could focus on optimising the algorithm for both speed and real-time RN generation, the latter becoming possible if commonplace adaptive optics were incorporated to stablise the system.

Next, we confirmed that the beams' COM measurements do follow the predicted normal distribution, with the standard deviations increasing as the turbulence strength increases, according to equation (\ref{eqn:ripsd}), (\ref{phasepsd}). The full RN generation and distribution scheme was then tested over a simulated free-space link of $10km$, with the resultant pair of 120,000-bit keys having a fidelity of $84\%$. This bodes well for potential applications. Two remote observers could then use their keys to secure a classical communication channel (which we demonstrated by encrypting/decrypting an image, figure \ref{fig:protocol}) or even use the extracted RNs to choose the random bases required in quantum communication protocols such as BB84 \cite{bennett2014quantum}. Finally, we investigated how the additional of simple, independent black boxes by both Alice and Bob to the setup effectively protects the channel from an eavesdropper who could potentially hack the entire free-space portion of the link.

It is important to compare any new proposal with existing protocols. Existing legacy free-space optical links typically employ techniques such as cryptographic hashing for digital signatures to initially authenticate both parties and thereafter use an asymmetric cryptography scheme to communicate. The protocol as outlined here would require a once-off authentication of the communication link between Alice and Bob. This is not unlike existing protocols which invariably require the two parties to be authenticated to, in the first instance, both be in possession of some mutual information for later use. However, once Alice and Bob have initially been authenticated in this scheme, they thereafter have no need to send private/public keys over the communication link to encrypt and decrypt further communication. They derive the same RNs remotely, and if we assume that an attacker only has access to the free-space portion of the link (i.e. the initial beam profiles and black boxes are secret and trusted, see figure \ref{fig:schematic}), said attacker cannot glean the same RNs. This is an advantage over a legacy channel employing commonplace symmetric or asymmetric cryptography schemes which generally necessitate the sending of private/public keys over the network. There is also potential for our protocol to instead be modified and added to existing secure communication links, thereby adding another layer of security, one which uses turbulence to its advantage rather than trying to overcome it like other free-space links.

Finally, it is worth mentioning that future developments of such a system may well incorporate classical communication capabilities in the free-space link itself so that the link serves a dual purpose: during periods of inadequate RN generation and distribution, the link could be used for ordinary optical communication; during periods of reasonable RN generation and distribution potential, it could be used for the purpose outlined above. For example, if a RN extraction algorithm is found to generate poor random numbers during times of very weak turbulence (when Alice and Bob’s beams are hardly distorted) then the system could be used for ordinary optical communication (by having one party encode information in, say, the beam's spatial profiles, and sending the beam to the other party) and not for generating random numbers. However, if the atmospheric turbulence then increases such that RN generation is more efficient and optical communication less efficient, the system could switch from optical communication to the RN generation scheme. The random numbers generated could then be used to either encrypt communication over a secondary link or encrypt later optical communication over the same link. This would require real-time measuring of the turbulence strength, which is possible with modern systems.

\section*{Acknowledgements}
\label{sec:acknowledgements}

The authors thank A. Vall\'es for his insightful discussion. N. B. acknowledges support from the CSIR DST-IBS programme.

\section*{References}

\bibliographystyle{unsrt}
\bibliography{References_File}

\end{document}